\documentclass{article}
\usepackage{amssymb}
\usepackage{amsmath}
\usepackage{latexsym}
\usepackage{bm}
\usepackage{graphicx}
\usepackage{lscape}
\usepackage{booktabs}
\usepackage{indentfirst}
\begin{document}

\title{Light (anti)nuclei production in Cu+Cu collisions at $\sqrt{s_{\rm{NN}}}=200$~GeV}
\author{Liu Feng-Xian$^{1,2}$, Chen Gang$^{1,2}$\footnote{Corresponding Author: chengang1@cug.edu.cn}, She Zhi-Lei$^{1,2}$,Dai-Mei Zhou$^3$, Xie Yi-Long$^2$}
\date{}
\maketitle
\begin{center}
\mbox{${^1}$Institute of Geophysics and Geomatics, China University of Geosciences, Wuhan 430074, China}\\
\mbox{${^2}$School of Mathematics and Physics, China University of Geosciences, Wuhan 430074, China}\\
\mbox{${^3}$ Institute of Particle Physics, Central China Normal University, Wuhan 430079, China}

\end{center}

\begin{abstract}
The production of light (anti)nuclei have been investigated using the dynamically constrained phase-space coalescence model based on the final-state hadrons generated by the PACIAE model in Cu+Cu collisions at $\sqrt{s_{\rm{NN}}}=200$~GeV with $|\eta|<0.5$ and $0<p_T<8$~GeV/c. The results show that there is a strong centrality dependence of yields of $\rm d$, $\rm\overline d$, $\rm ^3He$, $\rm^3\overline {He}$, $\rm ^4He$, and $\rm^4\overline {He}$, i.e., their yields decrease rapidly with the increase of centrality, and the greater the mass is, the greater the dependence shows; whereas their ratio of antinucleus to nucleus and coalescence parameter $B_A$ remain constant as the centrality increases. In addition, the yields of (anti)nuclei are strongly dependent on the mass of the (anti)nuclei, indicating that the (anti)nuclei produced have mass scaling properties in high-energy heavy-ion collisions. Our results are consistent with the STAR experimental data.


 \end{abstract}

\section{Introduction}

Since Dirac predicted the existence of negative energy states of electrons in 1928~\cite{Dirac}, the predicted antiprotons($\overline p$)~\cite{Cham} and antineutrons($\overline n$)~\cite{Cork} were observed in 1955 and 1956, followed by antideuterons($\rm \overline d$), antitritons($\rm ^3\overline H$), and antihelium-3($\rm ^3\overline {He}$) in scientific experiments~\cite{Mas,Dor,Vis,Ant}. The strange antinucleus - antihypertriton($\rm ^3_{\overline \Lambda}\overline H$), comprising an antiproton, an antineutron, and an antilambda hyperon, were discovered in 2010 by the STAR Collaboration at the Relativistic Heavy Ion Collider (RHIC) at the Brookhaven National Laboratory. The measured yields of $\rm ^3_{\Lambda}H$($\rm ^3_{\overline \Lambda}\overline H$) and $\rm ^3He(^3\overline {He})$ are similar, suggesting an equilibrium in coordinate and momentum space populations of up, down, and strange quarks and antiquarks, unlike the pattern observed at lower collision energies~\cite{BIA}. The antimatter helium-4 nucleus($\rm ^4\overline {He}$), also known as the anti-$\rm \alpha$($\rm \overline \alpha$), consists of two antiprotons and two antineutrons(baryon number $B=-4$). 18 counts were detected at the STAR experiment in $10^9$ recorded Au+Au collisions at centre-of-mass energies of 200~GeV and 62~GeV per nucleon-nucleon pair~\cite{Agak}.

On the other hand, the theoretical study of nuclei and antinuclei has been undertaken for many years. It usually includes two steps. First the nucleons and hyperons are calculated with some selected models, such as the transport models. Then the light (anti)nuclei are calculated by the phase-space coalescence model~\cite{Mat,LWC,Zhang} and/or the statistical model~\cite{Top,And}, etc.
In this paper, using the parton and hadron cascade model(PACIAE)~\cite{Sa} to simulate the production of (anti)nucleons($ p,\overline p,n,\overline n$) and (anti)hyperons($\rm\Lambda,\overline\Lambda$) in Cu+Cu collisions at $\sqrt{s_{\rm{NN}}}=200$~GeV, we compare them with experimental data from the STAR Collaboration~\cite{Agg,Aga,Abe} to fix the model parameters. Then, a dynamically constrained phase-space coalescence model(DCPC)~\cite{Yan,Chen1,Chen2} is used to study the production and properties of $\rm d(\overline d)$, $\rm ^3He(^3\overline {He})$, and $\rm ^4He(^4\overline {He})$. We expect that their yields in Cu+Cu collisions may provide some information about the nature of $\rm ^4He(^4\overline {He})$.

\section{Models}

The PACIAE~\cite{Sa} is based on PYTHIA 6.4 and designed for various nuclear collisions. In general, PACIAE has four main physics stages. The first stage is parton initiation, in which the nucleus-nucleus collision is decomposed into the nucleon-nucleon(NN) collisions according to the collision geometry and NN total cross section. The NN collisions will produce parton(gluons, quarks and antiquarks), and a new matter called the quark-gluon matter(QGM) is obtained. Then the next stage is the parton rescattering, the rescattering among partons in QGM is randomly considered by the $2 \rightarrow 2$ LO-pQCD parton-parton cross sections~\cite{Com}. The hadronization proceeds after the parton rescattering. The partonic matter can be hadronized by the Lund string fragmentation regime~\cite{TSj} and/or the phenomenological coalescence model~\cite{Sa}. The final stage is the hadron rescattering, the hadronic matter continues rescattering until the hadronic freeze-out~\cite{Sa}.

With the final-state particles produced by PACIAE model, we calculate the production of light (anti)nuclei and (anti)hypernuclei with the DCPC model. The DCPC model has been studied and used into several collision systems, such as Au+Au~\cite{Chen1,Chen2,Chen3,Dong}, Pb+Pb~\cite{She}, and $p+\overline p$ collisions~\cite{Sit} for several years. According to DCPC model, the yield of a single particle can be calculated using the following integral

\begin{equation}
Y_1=\int_{H\leqslant E} \frac{d\vec qd\vec p}{h^3},
\end{equation}
where $H$ and $E$ present the Hamiltonian and energy of the particle, respectively. Similarly, the yield of N particle cluster can be estimated by the integral
\begin{equation}
Y_N=\int ...\int_{H\leqslant E} \frac{d\vec q_1d\vec p_1...d\vec
q_Nd\vec p_N}{h^{3N}}. \label{phas}
\end{equation}
While this equation has to meet the constraint conditions as follows:
\begin{align}
m_0\leqslant m_{inv}\leqslant m_0+\Delta m;\\
q_{ij}\leqslant D_0(\hspace{0.2cm} i \neq j;j=1,2,...,N).
\end{align}
where
\begin{equation}
m_{inv}=\Big[(\sum_{i=1}^N E_i)^2-(\sum_{i=1}^N p_i)^2\Big]^{1/2}
\end{equation}
$E_i$ and $p_i$ (i=1, 2, ... , N) denote the energies and momenta of particles. $m_0$ and $\Delta m$ respectively represent the rest mass and the allowed mass uncertainty. $D_0$ stands for the diameter of (anti)nuclei, and $q_{ij}=|\vec q_i-\vec q_j|$ presents the vector distance between particle  $i$ and particle $j$. Here, the diameters of the (anti)nuclei and (anti)hypernuclei are calculated by $D_0=2r_0 A^{1/3}$, $D_0=1.50, 2.02, 2.23$ fm for $\rm d(\overline d)$, $\rm ^3He(^3\overline {He})$, and $\rm ^4He(^4\overline {He})$ in the model, respectively~\cite{Sa,TAA,Ham,Nem}.  The integral over continuous distributions in Eq.(2) should be replaced by the sum over discrete distributions as the hadron position and momentum distributions from transport model simulation are discrete.

\section{Results}

First we produce the final-state particles using the PACIAE model. In the PACIAE simulations, we assume that the hyperons heavier than $\Lambda$ have already decayed before the creation of hypernuclei. We use the default values of model parameters given in the PYTHIA in our model, except the \textit{K} factor and the parameters parj(1), parj(2), and parj(3)(here, parj(1)=0.18, parj(2)=0.43, and parj(3)=0.40) were roughly fitted to the STAR data~\cite{Agg,Aga,Abe} in Cu+Cu collisions at $\sqrt{s_{\rm{NN}}}=200$~GeV for different centrality bins of 0-10\%, 10-20\%, 20-30\%, 30-40\%, and 40-60\%, in which, the yields of particles were calculated with $|\eta|<0.1$ and $0.4<p_T<1.2$~GeV/c for $p$ and $\overline p$, and $|\eta|<0.5$ and $0<p_T<8$~GeV/c for $\Lambda$ and $\overline\Lambda$. From Table 1, we can find the yields of final-state hadrons($ p,\overline p,\rm\Lambda,\overline\Lambda$) agree with STAR data within uncertainties.

\vspace{-0.5cm}
\begin{table*}[hp]
\small
\caption{Yields of particles($p, \overline p, \Lambda, \overline \Lambda $) in Cu+Cu collisions at $\sqrt{s_{\rm{NN}}}=200$~GeV
for different centrality bins, compared with the STAR data~\cite{Agg,Aga}.}
\setlength{\tabcolsep}{5.2pt}
\renewcommand{\arraystretch}{1.}
\vspace{-0.2cm}
\begin{center}
\begin{tabular}{ccccccc}\hline
Centrality & &$0-10\%$ & $10-20\%$ & $20-30\%$ & $30-40\%$ & $40-60\%$ \\\hline
$p$ &PACIAE &5.73$\pm$0.22 &4.15$\pm$0.25 &2.94$\pm$0.20 &2.07$\pm$0.18 &1.08$\pm$0.12 \\
$ $ &STAR  &5.56$\pm$0.02$\pm$0.19 &3.94$\pm$0.01$\pm$0.14 &2.76$\pm$0.01$\pm$0.10 &1.92$\pm$0.01$\pm$0.07 &1.09$\pm$0.02$\pm$0.05 \\
$\overline p$ &PACIAE &4.69$\pm$0.18 &3.41$\pm$0.21 &2.41$\pm$0.17&1.69$\pm$0.14 &0.88$\pm$0.09 \\
$ $ &STAR &4.54$\pm$0.02$\pm$0.13 &3.26$\pm$0.01$\pm$0.09 &2.32$\pm$0.01$\pm$0.06 &1.63$\pm$0.01$\pm$0.05 &0.93$\pm$0.02$\pm$0.04 \\
$\Lambda$ &PACIAE &4.24$\pm$0.13 &3.03$\pm$0.18 &2.15$\pm$0.15 &1.49$\pm$0.16 &0.78$\pm$0.10 \\
$ $ &STAR &4.68$\pm$0.45 &3.20$\pm$0.31 &2.13$\pm$0.21 &1.40$\pm$0.14 &0.72$\pm$0.07 \\
$\overline \Lambda$ &PACIAE &3.64$\pm$0.11 &2.64$\pm$0.18 &1.88$\pm$0.15 &1.32$\pm$0.13 &0.70$\pm$0.08 \\
$ $ &STAR &3.79$\pm$0.37 &2.60$\pm$0.25 &1.75$\pm$0.17 &1.16$\pm$0.11 &0.60$\pm$0.06 \\\hline
\end{tabular}
\end{center}
\end{table*}

Then we use the nucleons and hyperons produced within PACIAE as the input of DCPC model to generate 100 million minimum bias events for Cu+Cu collisions at $\sqrt{s_{\rm{NN}}}=200$~GeV and obtain the integrated yields dN/dy of $\rm d(\overline d)$, $\rm ^3He(^3\overline {He})$, and $\rm ^4He(^4\overline {He})$ for different centrality classes of 0-10\%, 10-20\%, 20-30\%, 30-40\%, and 40-60\%, as shown in Table 2 and plotted in Fig. 1. It is obvious that the integrated yields dN/dy of light nuclei and light antinuclei calculated by the DCPC model decrease with the increase of centrality. The yields of antinuclei are less than those of their corresponding nuclei, and the greater the mass of (anti)nuclei is, the lower the yield is. One sees in the Fig. 1 that the PACIAE+DCPC results
agree well with the experimental data.

\begin{figure}[hp]
\begin{center}
\includegraphics[width=15cm,height=9cm]{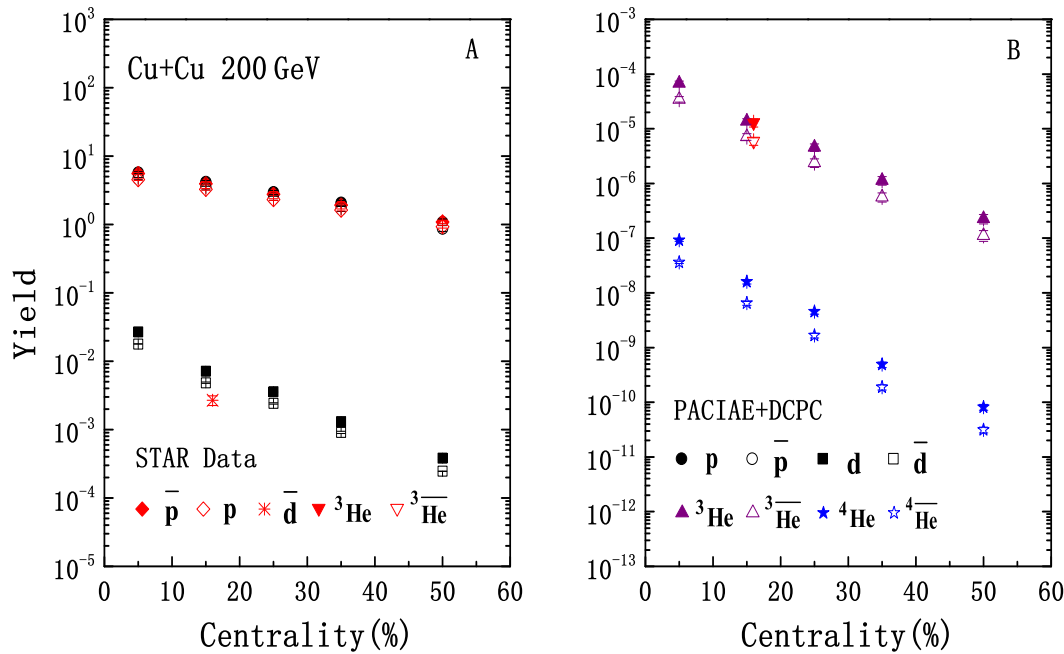}
\caption{The integrated yield of particles for $p$, $\overline p$, $\rm d$, $\rm\overline d$, $\rm ^3He$, $\rm^3\overline {He}$, $\rm ^4He$, and $\rm^4\overline {He}$ calculated by PACIAE+DCPC model in Cu+Cu collisions at $\sqrt{s_{\rm{NN}}}=200$~GeV , as a function of centrality. The data are from STAR~\cite{Agg,Zhou}.}
\end{center}
\end{figure}

\vspace{-0.5cm}
\begin{table*}[hp]
\small
\caption{The integrated yield of particles for $\rm d$, $\rm\overline d$, $\rm ^3He$, $\rm^3\overline {He}$, $\rm ^4He$, and $\rm^4\overline {He}$ in Cu+Cu collisions of $\sqrt{s_{\rm{NN}}}=200$~GeV, calculated by PACIAE+DCPC model for various centrality classes. The STAR data for minimum bias are taken from~\cite{Zhou}.}
\setlength{\tabcolsep}{2.6pt}
\renewcommand{\arraystretch}{1.2}
\vspace{-0.2cm}
\begin{center}
\begin{tabular}{ccccccc}\hline
Nucleus &STAR& $0-10\%$ & $10-20\%$ & $20-30\%$ & $30-40\%$ & $40-60\%$ \\\hline
$\rm d^a$ & &(26.77$\pm$0.15)E-03 &(7.14$\pm$0.15)E-03 &(3.57$\pm$0.10)E-03 &(1.31$\pm$0.05)E-03 &(3.84$\pm$0.15)E-04 \\
$\rm\overline d^a$ &(2.69$\pm$0.43)E-03&(17.90$\pm$0.08)E-03 &(4.85$\pm$0.11)E-03 &(2.42$\pm$0.06)E-03 &(9.08$\pm$0.34)E-04 &(2.46$\pm$0.02)E-04 \\
$\rm ^3{He}^b$ &(1.29$\pm$0.22)E-05&(6.76$\pm$0.66)E-05 &(1.37$\pm$0.16)E-05 &(4.58$\pm$0.71)E-06 &(1.13$\pm$0.20)E-06 &(2.26$\pm$0.45)E-07 \\
$\rm ^3\overline {He}^b$ &(0.59$\pm$0.09)E-05&(3.46$\pm$0.40)E-05 &(7.13$\pm$0.95)E-06 &(2.38$\pm$0.43)E-06 &(5.62$\pm$1.06)E-07 &(1.12$\pm$0.25)E-07 \\
$\rm ^4{He}^c$ & &(9.16$\pm$0.44)E-08 &(1.59$\pm$0.13)E-08 &(4.53$\pm$0.46)E-09 &(4.88$\pm$0.58)E-10 &(8.13$\pm$0.98)E-11 \\
$\rm ^4\overline {He}^c$ & &(3.60$\pm$0.18)E-08 &(6.47$\pm$0.52)E-09 &(1.66$\pm$0.17)E-09 &(1.88$\pm$0.21)E-10 &(3.12$\pm$0.38)E-11 \\\hline
\multicolumn{5}{l}{$^a$ calculated with $\Delta m=0.89$~MeV for $d$, $\overline d$;}\\
\multicolumn{5}{l}{$^b$ calculated with $\Delta m=1.58$~MeV for $\rm{^3{He}}$, $\rm{^3{\overline{He}}}$;}\\
\multicolumn{5}{l}{$^c$ calculated with $\Delta m=1.85$~MeV for $\rm{^4{He}}$, $\rm{^4{\overline{He}}}$.}\\
\end{tabular}
\end{center}
\end{table*}

We also calculated the ratios of light antinuclei to light nuclei (${\rm\overline d/d}$, $\rm {^3\overline {He}/^3{He}}$, and $\rm {^4\overline{He}/^4{He}}$), as well as their mixing ratios of ${\rm d}/p$, ${\rm\overline d}/{\overline p}$, $\rm ^3{He}/d$, ${\rm ^3\overline {He}/{\overline d}}$, $\rm ^4{He}/\rm ^3{He}$, and ${\rm ^4\overline {He}}/{\rm ^3\overline {He}}$ in different centrality Cu+Cu collisions of $\sqrt{s_{\rm{NN}}}=200$~GeV, presented in Table 3 and plotted in Fig. 2. For comparison, experimental results from STAR are also given within the Table 3 and Fig. 2. We can see, in the upper section of Table 3 and Fig. 4A, that the yield ratios of light antinuclei to light nuclei from central to peripheral collisions remain unchanged, although their yields decrease rapidly with the centrality as shown in Table 2 and Fig. 1. Their values fluctuate around 0.66 for ${\rm\overline d/d}$, 0.50 for $\rm {^3\overline {He}/^3{He}}$, and 0.38 for $\rm {^4\overline{He}/^4{He}}$ , indicating the greater the mass of (anti)nuclei is, the more difficult it is to produce an antinucleus than the corresponding nucleus. It can be seen from Fig. 4B and the lower section of Table 3 that the mixing ratios of heavier to lighter (anti)nuclei from central to peripheral collisions decreases, suggesting that it is easier to produce light nuclei in the central collision region; and the heavier the (anti)nucleus is, the smaller the mixing ratio is, showing that the heavier nucleus is more difficult to produce than that of the light nucleus.

\begin{table*}[hp]
\small
\caption{The ratios of particles in Cu+Cu collisions of $\sqrt{s_{\rm{NN}}}=200$~GeV calculated by PACIAE+DCPC model for various centrality classes, compared with experimental data from STAR~\cite{Agg,Zhou}.}
\setlength{\tabcolsep}{2.3pt}
\renewcommand{\arraystretch}{1.2}
\vspace{-0.2cm}
\begin{center}
\begin{tabular}{ccccccc}\hline
Centrality &STAR& $0-10\%$ & $10-20\%$ & $20-30\%$ & $30-40\%$ & $40-60\%$ \\\hline
$\rm \overline d/d$  & &0.67$\pm$0.05 &0.68$\pm$0.03 &0.68$\pm$0.03 &0.69$\pm$0.04 &0.64$\pm$0.03 \\
$\rm ^3\overline {He}/^3{He}$  &0.46$\pm$0.10 &0.51$\pm$0.08 &0.52$\pm$0.09 &0.52$\pm$0.13 &0.50$\pm$0.13 &0.50$\pm$0.15 \\
$\rm ^4\overline {He}/^4{He}$  &&0.39$\pm$0.03 &0.41$\pm$0.05 &0.37$\pm$0.06 &0.39$\pm$0.07 &0.38$\pm$0.07 \\
\hline
${\rm d}/p$   &&(4.67$\pm$0.18)E-03 &(1.72$\pm$0.11)E-03 &(1.21$\pm$0.09)E-03 &(6.33$\pm$0.60)E-04 &(3.56$\pm$0.42)E-04 \\
${\rm\overline d}/{\overline p}$ &(1.19$\pm$0.19)E-03&(3.82$\pm$0.15)E-03 &(1.42$\pm$0.10)E-03 &(1.01$\pm$0.08)E-03 &(5.37$\pm$0.49)E-04 &(2.80$\pm$0.29)E-04 \\
$\rm ^3{He}/d$  & &(2.53$\pm$0.25)E-03 &(1.92$\pm$0.23)E-03 &(1.28$\pm$0.21)E-03 &(8.62$\pm$1.57)E-04 &(5.89$\pm$1.20)E-04 \\
${\rm ^3\overline {He}/{\overline d}}$  &(2.19$\pm$0.49)E-03&(1.93$\pm$0.23)E-03 &(1.47$\pm$0.20)E-03 &(9.83$\pm$1.80)E-04 &(6.19$\pm$1.19)E-04 &(4.55$\pm$1.02)E-04 \\
$\rm ^4{He}/\rm ^3{He}$ &&(1.36$\pm$0.15)E-03 &(1.16$\pm$0.17)E-03 &(9.89$\pm$1.84)E-04 &(4.32$\pm$0.93)E-04 &(3.59$\pm$0.84)E-04 \\
${\rm ^4\overline {He}}/{\rm ^3\overline {He}}$  & &(1.04$\pm$0.14)E-03 &(9.07$\pm$1.42)E-04 &(6.97$\pm$1.45)E-04 &(3.35$\pm$0.74)E-04 &(2.79$\pm$0.71)E-04 \\\hline
\end{tabular}
\end{center}
\end{table*}

\begin{figure*}[htbp]
\begin{center}
\includegraphics[width=15cm,height=9cm]{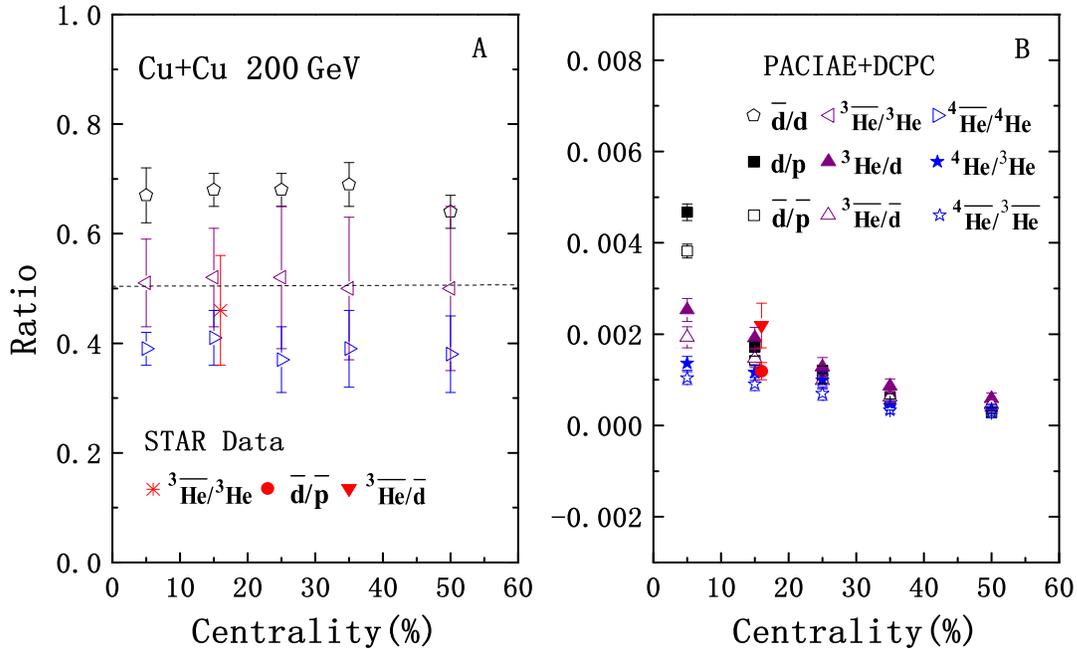}
\caption{A. The integrated yield ratio of antinucleus to nucleus ($\rm \overline d/d$, $\rm ^3\overline {He}/^3{He}$, $\rm ^4\overline {He}/^4{He}$); B. The mixing ratio of light (anti)nuclei ($\rm d/p$, $\rm ^3{He}/d$,
$\rm ^4{He}/^3{He}$, $\rm \overline d/{\overline p}$, $\rm ^3\overline {He}/{\overline d}$, $\rm ^4\overline {He}/{^3\overline {He}}$), calculated by PACIAE+DCPC model in Cu+Cu collisions at $\sqrt{s_{\rm{NN}}}=200$~GeV, as a function of centrality. The data are from STAR~\cite{Agg,Zhou}.}
\end{center}
\end{figure*}

Meanwhile, we analyse the distribution of integrated yields dN/dy of nuclei ($p$, $\rm d$, $\rm ^3He$, and $\rm ^4He$), and their antimatters ($\overline p$, $\rm\overline d$, $\rm^3\overline {He}$, and $\rm^4\overline {He}$) with the mass number $A$ in Cu+Cu collisions at $\sqrt{s_{\rm{NN}}}=200$~GeV for three different centrality bins of 0-10\%, 10-30\%, and 30-60\%, as shown in Fig. 3, respectively. From the comparison between the calculation by PACIAE+DCPC model(open symbols) and the STAR data~\cite{Agg,Aga,Zhou}(solid symbols), our results are consistent with the STAR measurement within uncertainties, and one can easily find that the integrated yields of (anti)nuclei all decrease rapidly with the increase of mass number, which exhibit exponential behaviour as a function of mass number. Moreover, in central 200~GeV Au+Au collisions, the STAR Collaboration~\cite{Agak} have observed an exponential yield consistent with expectations from thermodynamic model~\cite{Arm,Bra} and coalescent nucleosynthesis model~\cite{Sato}. In addition, according to Refs.~\cite{And,Arm,Liu}, antimatter nuclei with baryon number $B<-1$ have been observed only as rare products of interactions at particle accelerators, where the rate of antinucleus production in high-energy collisions decreases by a factor of about 1,000 with each additional antinucleon. Our results calculated from PACIAE+DCPC model can also prove this, as shown in Table 2 and Fig. 3. So this behaviour can provides a rough estimate of the production for heavier (anti)nuclei. In this figure, the curve is fitted to the data point using an equation as ~\cite{25,31,32}:

\begin{equation}
E_{A}\frac{d^3N_{A}}{d^3P_{A}} \propto e^{-Am_p/T}.
\end{equation}
where $E_A\frac{d^3N}{d^3P}$ stands for the invariant yield of (anti)nuclei, $P_A$ is the momentum of
(anti)nuclei, $T$ is the temperature at hadronic freeze-out, and $m_p \simeq m_n$ is the mass of proton and neutron.

\begin{figure*}[htbp]
\begin{center}
\includegraphics[width=11cm,height=9cm]{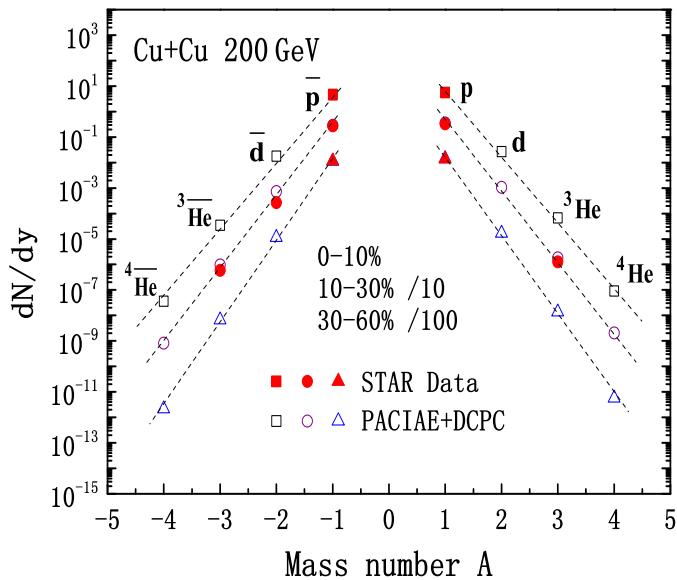}
\caption{Atomic mass number $A$ dependence of the integrated yield dN/dy of light (anti)nuclei in Cu+Cu collisions at $\sqrt{s_{\rm{NN}}}=200$~GeV for three different centrality bins of 0-10\%, 10-30\%, and 30-60\%. The solid symbols represent the STAR data~\cite{Agg,Zhou}, while the open ones are calculated by PACIAE+DCPC model. The lines represent the model result's exponential fit for the positive matters(right) and negative matters(left) with formula $e^{-Am_p/T}$.}
\end{center}
\end{figure*}

The yield per participant nucleon may reflects the formation probability of a hadron from the bulk. We define a relative yield $R_{CY}(N_{part})$ as a measure of the dependence of the (anti)nuclei on the collision system¡¯s size and density,

\begin{equation}
R_{CY}(N_{part})=\frac{(dN/dy)/N_{part}}{[(dN/dy)/N_{part}]^{Peripheral}}.
\end{equation}

Figure 4 shows the relative yields $R_{CY}(N_{part})$ of $p$, $\overline p$, $\Lambda$, $\overline \Lambda$, $\rm d$, $\rm\overline d$, $\rm ^3He$, $\rm^3\overline {He}$, $\rm ^4He$, and $\rm^4\overline {He}$ calculated by PACIAE+DCPC model in Cu+Cu collisions at $\sqrt{s_{\rm{NN}}}=200$~GeV. The results are normalized by peripheral collisions (40-60\%). We find that the yields of light (anti)nuclei per participant nucleon increase rapidly with the increase of the number of $N_{part}$ as the $N_{part} > 60$. This distribution properties of light nuclei and light antinuclei production in Cu+Cu collisions at $\sqrt {s_{NN}} = 200$ GeV depend on their mass number, i.e., the greater the mass number is, the faster the yield increases. Using this same model in Au+Au collisions~\cite{Chen3} and Pb+Pb collisions~\cite{She}, the relative yields $R_{CY}(N_{part})$ show the same trend.

\begin{figure*}[htbp]
\begin{center}
\includegraphics[width=11cm,height=8cm]{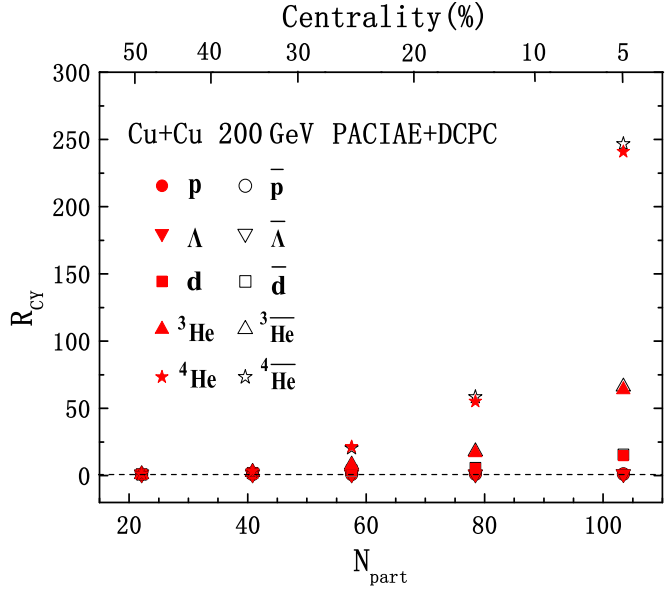}
\caption{The integrated yield $dN/dy$ at midrapidity for $p$, $\overline p$, $\Lambda$, $\overline \Lambda$, $\rm d$, $\rm\overline d$, $\rm ^3He$, $\rm^3\overline {He}$, $\rm ^4He$, and $\rm^4\overline {He}$ divided by $N_{part}$, normalized to the peripheral collisions (40-60\%), plotted as a function of $N_{part}$. The results are calculated by the PACIAE+DCPC model in Cu+Cu collisions at $\sqrt{s_{\rm{NN}}}=200$~GeV. The solid and open symbols
represent the positive and negative nuclei, respectively.}
\end{center}
\end{figure*}

\begin{figure*}[htbp]
\begin{center}
\includegraphics[width=15cm,height=9cm]{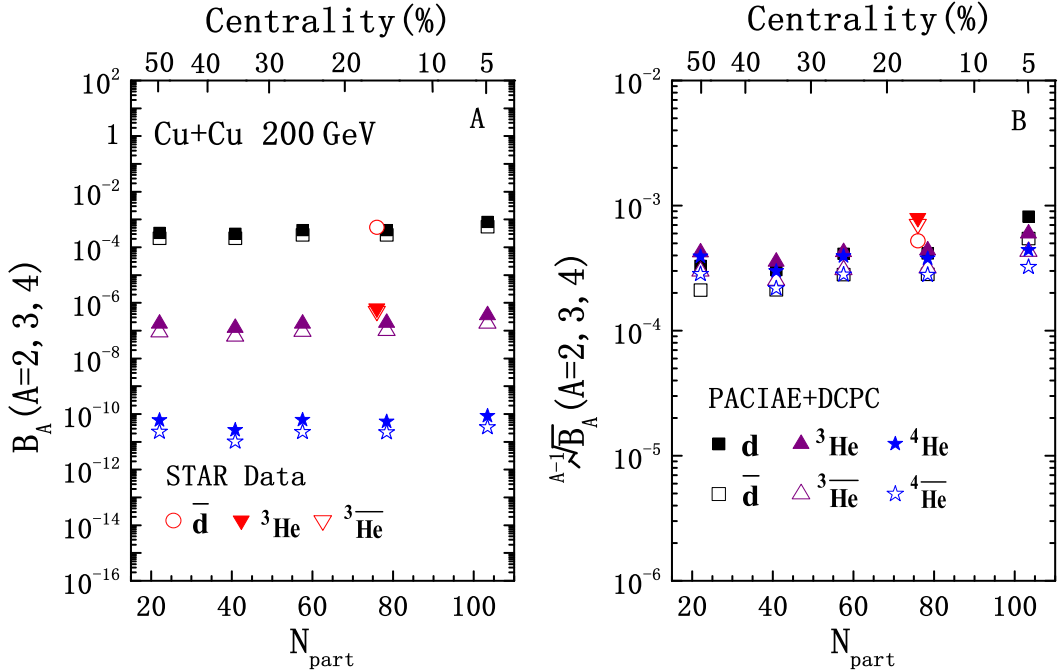}
\caption{A. Coalescence parameter $B_{A}$ (on left), B. Coalescence parameter $\sqrt[A-1]{B_{A}}$ (on right), as a function of $N_{part}$ for (anti)nuclei in Cu+Cu collisions at $\sqrt{s_{\rm{NN}}}=200$~GeV. The solid and open symbols represent the positive and negative nuclei, respectively. The data are from STAR~\cite{Zhou}.}
\end{center}
\end{figure*}

In heavy-ion collisions, the coalescence process of light (anti)nuclei, and (anti)hypernuclei is historically described by the coalescence parameter $B_A$. The differential invariant yield for the production of (anti)nuclei is related~\cite{RSC,HH} to the primordial yields of nucleons by

\begin{equation}
E_{A}\frac{d^3N_{A}}{d^3P_{A}}=B_{A}(E_{P}\frac{d^3N_{P}}{d^3P_{P}})^Z(E_{n}\frac{d^3N_{n}}{d^3P_{n}})^{A-Z}\approx B_{A}(E_{P}\frac{d^3N_{P}}{d^3P_{P}})^A.
\end{equation}
where $N_{A}$, $N_{p}$, and $N_{n}$ denote the number of the (anti)nuclei, their constituent (anti)protons and (anti)neutrons, respectively; $A$ and $Z$ are the atomic mass number and atomic number, respectively; $E_A\frac{d^3N}{d^3P}$ stands for the invariant yield of (anti)nucleons or (anti)nuclei; $P_{p}$ and $P_{A}$ are their momenta, where $P_{A}= AP_{p}$. $B_{A}$ is the coalescence parameter related to the freeze-out correlation volume~\cite{RSC,HH},

\begin{equation}
B_{A}\propto V_{f}^{1-A}.
\end{equation}

Figure 5 presents the distribution of $B_{A}$ as a function of $N_{part}$. $B_{2}$, $B_{3}$, and $B_{4}$ calculated based on the invariant yields of $\rm d(\overline d)$, $\rm ^3He(^3\overline {He})$, and $\rm ^4He(^4\overline {He})$. $B_{2}\propto 1/V_{f}$, $B_{3}\propto 1/(V_{f})^2$, and $B_{4}\propto 1/(V_{f})^3$, according to the Eq.(9). $B_{A}$($A=2,3,4$) remains roughly unchanged from central to peripheral collisions by using PACIAE+DCPC model as shown in Fig. 5, and the positive nuclei are a little bigger than the negative nuclei. The $B_{A}$ of $\rm ^4He(^4\overline {He})$ is smaller than one of $\rm ^3He(^3\overline {He})$, which is smaller than $\rm d(\overline d)$, showing that combining to produce a heavier nucleus is harder than producing a lighter one. The results obtained from our model are also in agreement with the experimental data from STAR~\cite{Zhou} within error ranges.


In this paper we have employed the DCPC model to investigate the light (anti)nuclei production and the centrality dependence based on the final hadronic state generated by the PACIAE model in Cu+Cu collisions at $\sqrt{s_{\rm{NN}}}=200$~GeV with $|\eta|<0.5$ and $0 < p_T < 8$~GeV/c. The results show that the yields of $\rm d(\overline d)$, $\rm ^3He(^3\overline {He})$, and $\rm ^4He(^4\overline {He})$ decrease rapidly with the increase of centrality. And the integrated yields of (anti)nuclei all decrease rapidly with the increase of mass number, which exhibit exponential behaviour. However, the yield ratios of light antinuclei ($\rm \overline d$, $\rm ^3\overline {He}$, and $\rm ^4\overline {He}$) to light nuclei ($\rm d$, $\rm ^3He$, and $\rm ^4He$) are independent on centrality, but the mixing ratios of light (anti)nuclei (${\rm d}/p$, $\rm ^3{He}/d$,$\rm ^4{He}/^3{He}$, ${\rm\overline d}/{\overline p}$, $\rm ^3\overline {He}/{\overline d}$, and $\rm ^4\overline {He}/{^3\overline {He}}$) are  dependent on centrality.

In addition, we researched the relative yields $R_{CY}(N_{part})$ per $N_{part}$ of light (anti)nuclei, normalized to the values obtained in the peripheral collisions (40-60\%). It is found that the yields of light (anti)nuclei per participant nucleon increase with $N_{part}$ as $N_{part}>60$, and the yields of heavy nuclei increase more rapidly than that of light nuclei. Obviously, this distribution properties of light antinuclei production in Cu+Cu collisions at $\sqrt{s_{\rm{NN}}}=200$~GeV depend on their mass.
At last, we also discussed coalescence parameter $B_A$ to measure the difficulty in synthesizing nucleus. We find that coalescence parameter $B_A$ remains roughly unchanged from central to peripheral collisions, but it depends on their mass, i.e., producing a heavier nucleus is harder than producing a lighter one.
Our model results are also consistent with the STAR data.
The consistency between our model results and the corresponding experimental data demonstrates that the PACIAE+DCPC model is able to describe the production of light (anti)nuclei in the relativistic heavy-ion collisions.

\begin{center} {ACKNOWLEDGEMENT} \end{center}

The financial support from NSFC(11475149) is acknowledged, and supported by the high-performance computing
platform of China University of Geosciences. The authors thank for helpful discussions.

\end{document}